\documentclass[sigconf]{acmart}
\AtBeginDocument{%
  }

\usepackage{siunitx}
\copyrightyear{2025}
\acmYear{2025}
\setcopyright{rightsretained}
\acmConference[TEI '25]{Nineteenth International Conference on Tangible,
Embedded, and Embodied Interaction}{March 4--7, 2025}{Bordeaux / Talence,
France}
\acmBooktitle{Nineteenth International Conference on Tangible, Embedded, and
Embodied Interaction (TEI '25), March 4--7, 2025, Bordeaux / Talence, France}\acmDOI{10.1145/3689050.3705973}
\acmISBN{979-8-4007-1197-8/25/03}

\begin{document}

\title[Plug-n-play e-knit: prototyping large-area e-textiles using machine-knitted magnetically-repositionable sensor network]{Plug-n-play e-knit: prototyping large-area e-textiles using machine-knitted magnetically-repositionable sensor networks}


\author{Yifan Li}
\affiliation{
  \institution{The University of Tokyo}
  \country{Japan}}
\email{yifan217@akg.t.u-tokyo.ac.jp}

\author{Ryo Takahashi}
\affiliation{
  \institution{The University of Tokyo}
  \country{Japan}}
\email{takahashi@akg.t.u-tokyo.ac.jp}

\author{Wakako Yukita}
\affiliation{
  \institution{The University of Tokyo}
  \country{Japan}}
\email{yukita@bhe.t.u-tokyo.ac.jp}

\author{Kanata Matsutani}
\affiliation{
 \institution{Nagoya Institute of Technology}
 \country{Japan}}
\email{k.matsutani.640@stn.nitech.ac.jp}

\author{Cedric Caremel}
\affiliation{
  \institution{The University of Tokyo}
  \country{Japan}}
\email{cedric@akg.t.u-tokyo.ac.jp}

\author{Yuhiro Iwamoto}
\affiliation{
 \institution{Nagoya Institute of Technology}
 \country{Japan}}
\email{iwamoto.yuhiro@nitech.ac.jp}

\author{Sunghoon Lee}
\affiliation{
  \institution{The University of Tokyo}
  \country{Japan}}
\email{sunghoon@ntech.t.u-tokyo.ac.jp}

\author{Tomoyuki Yokota}
\affiliation{
  \institution{The University of Tokyo}
  \country{Japan}}
\email{yokota@ntech.t.u-tokyo.ac.jp}

\author{Takao Someya}
\affiliation{
  \institution{The University of Tokyo}
  \country{Japan}}
\email{someya@ee.t.u-tokyo.ac.jp}

\author{Yoshihiro Kawahara}
\affiliation{
  \institution{The University of Tokyo}
  \country{Japan}}
\email{kawahara@akg.t.u-tokyo.ac.jp}

\renewcommand{\shortauthors}{Yifan Li, Ryo Takahashi et al.}

\begin{abstract}

Prototyping electronic textile involves embedding electronic components into fabrics to develop smart clothing with specific functionalities.
However, this process is still challenging since the complicated wiring setup is required during experimental phases.
This paper presents plug-n-play e-knit, a large-scale, repositionable e-textile for providing trial-and-error prototyping platforms across the textile.
Plug-n-play e-knit leverages industrial digital knitting machines loaded with conductive thread to automatically embed a communication and power supply network into garments, in addition to using soft magnet connectors to rearrange electronic components while preserving the stretchability of the garment.
These combinations enable users to quickly establish e-textile sensor networks, and moreover test the performance and optimal placement of the electric devices on the textile.
We demonstrated that our textiles leveraging custom  I$^2$C protocols could achieve the motion-resilient motion-tracking sensor network over a \SI{2700}{cm^2} garment area.

\end{abstract}

\ccsdesc[500]{Human-centered computing~Interaction devices}

\keywords{E-textiles, e-knit, PME, wearable, prototyping tool, smart clothing}

\begin{teaserfigure}
  \includegraphics[width=1.0\textwidth]{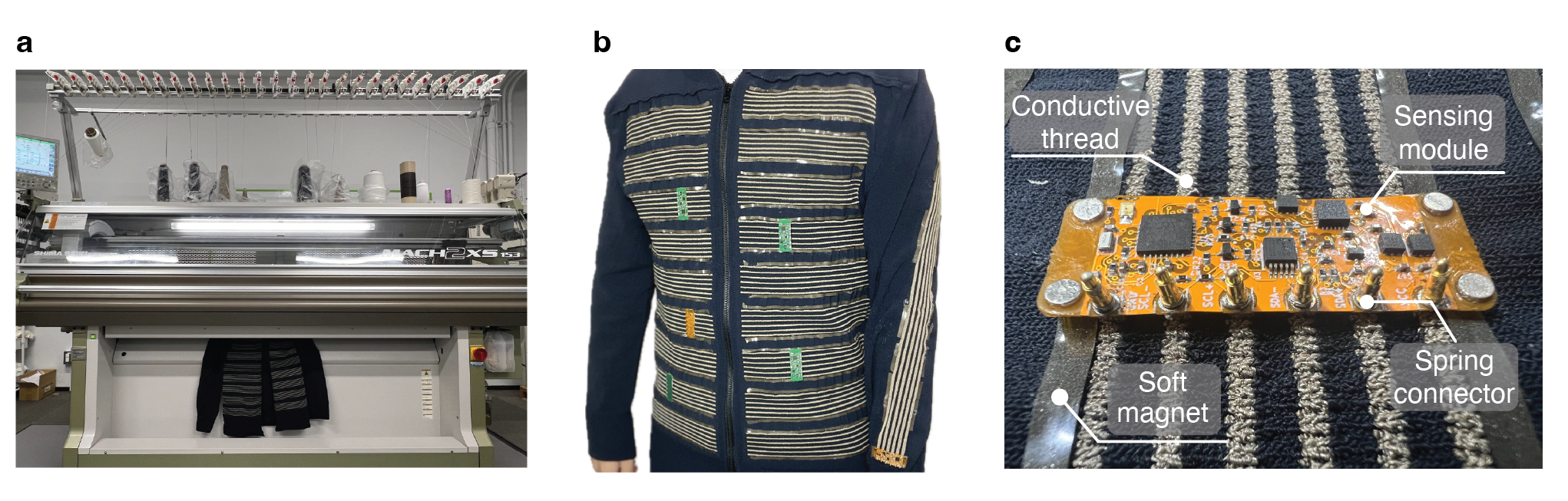}
  \caption{Overview of plug-n-play e-knit. Plug-n-play e-knit is a (a) machine-knitted e-textile prototyping tool, which enables (b) large-area, reconfigurable and on-textile sensing. (c) It also provides a non-invasive sensor connection method by using soft magnet.}
  \Description{}
  \label{fig:teaser}
\end{teaserfigure}

\maketitle

\section{INTRODUCTION}

Developing electronic textile (E-textile) involves repeated disassembly of integrated electronic devices to find the optimal device position in the field of activity recognition and healthcare~\cite{ohiri_e-textile_2022, griffin_analysis_2016,hill_threadboard_2021}.
For example, monitoring lower-limb rehabilitation requires the precise placement of sensors at different patients' injured joints and muscles~\cite{woelfle_plug-and-play_2023}.
E-textile prototyping tools provide means to quickly test and evaluate the proof-of-concept e-textiles~\cite{stanley_modular_2022, han_towards_2021, tseghai2020review}. 
However, current e-textile prototyping tools use invasive connection methods such as stitching~\cite{buechley_lilypad_2008}, welding~\cite{fromme_metal-textile_2021}, adhesives~\cite{ohiri_e-textile_2022}, causing the textile damage when connecting the sensing module with the e-textiles.
The lack of a method for integrating large-area conductive threads restricts the possible installation positions of the sensing module, preventing it from covering a large portion of the body~\cite{jones_wearable_2020, woelfle_plug-and-play_2023}.

To address the issues, we propose plug-n-play e-knit, a large-scale, reconfigurable, scalable e-textile prototyping tool. 
The plug-n-play e-knit design includes 1) the machine-knitted textile-based communication and power supply network for sensing modules on the textiles and 2) the soft magnet connector to rearrange these modules to the textile.
First, the knitting machine directly integrates stretchable communication and power supply network into the garment at scale, enabling to freely place the modules across the textile.
Then, the soft magnet connector compromised of Permanent Magnet Elastomer (PME)~\cite{Shembekar_PME_2021} offers repeated replacement of the sensing modules without damaging the textile and hindering the user motions.
Sensing modules based on customized  I$^2$C protocols can function by simply touching the corresponding position of the conductive threads.
With these configurations, plug-n-play e-knit provides an e-textile platform that allows repeatable on-textile sensor performance tests and e-textile application development such as body motion tracking.

\section{RELATED WORK}

\subsection{E-textiles}

Since daily clothing covers a large area of the human body, the integration of electric functionality into the clothing enables the monitoring of comprehensive physiological and environmental signals across the body.
Prior research has proposed large-scale, highly integrated e-textiles using two types of connectivity: wired approach and wireless approach.
As for the wired approach, researchers typically incorporate conductive threads into textiles via weaving~\cite{wicaksono_tailored_2020}, knitting~\cite{song2010knit}, and embroidery~\cite{hamdan2018embridery, Wang2023Emtex}, subsequently connecting them to sensor boards through welding or stitching~\cite{linz2005stitching}. Wired e-textiles typically exhibit higher energy efficiency~\cite{stanley_review_2022}.
By contrast, the wireless approach does not necessitate the direct integration of electronic components into textiles. 
To achieve non-contact communication and power transfer, researchers incorporate conductive coils into textiles by sewing~\cite{fobelets2019wirelessknit} or printing~\cite{li2018printcoil}.
These coils act as inductors and form an inductive coupling or Near Field Communication (NFC) terminal~\cite{lin_digitally-embroidered_2022, takahashi2021meander, takahashi2022meander,takahashi2018cuttable}, enabling electronic devices placed near them to receive power and exchange data.
However, achieving large-scale, repositionable, and customizable e-textiles remains highly challenging because of transmission noise~\cite{noda_I2We_2019}, the lack of lightweight, and non-invasive connectors~\cite{stanley_review_2022}.

\subsection{Plug-n-Play Wearables}

In prototyping e-textiles, the plug-and-play feature facilitates the easy attachment and removal of electronic components, which not only reduces the debugging time for developers but also minimizes the damage to the textiles caused by component disassembly.
There are various types of the plug-n-play connections including primarily employs snap fasteners~\cite{lesnikowski_research_2016, jones_swatch-bits_2020}, pogo pins and magnets~\cite{socks}, pin headers~\cite{markvicka_STEAM_2018}, conductive hooks~\cite{woelfle_plug-and-play_2023}, and adhesives~\cite{stanley_modular_2022}. 
However, these methods have issues such as poor stretchability, causing damage to textiles, and inconveniences in changing the positions of sensing modules~\cite{stanley_review_2022}. 

\section{PLUG-N-PLAY E-KNIT DESIGN}
\begin{figure}[t!]
  \centering
  \includegraphics[width=0.45\textwidth]{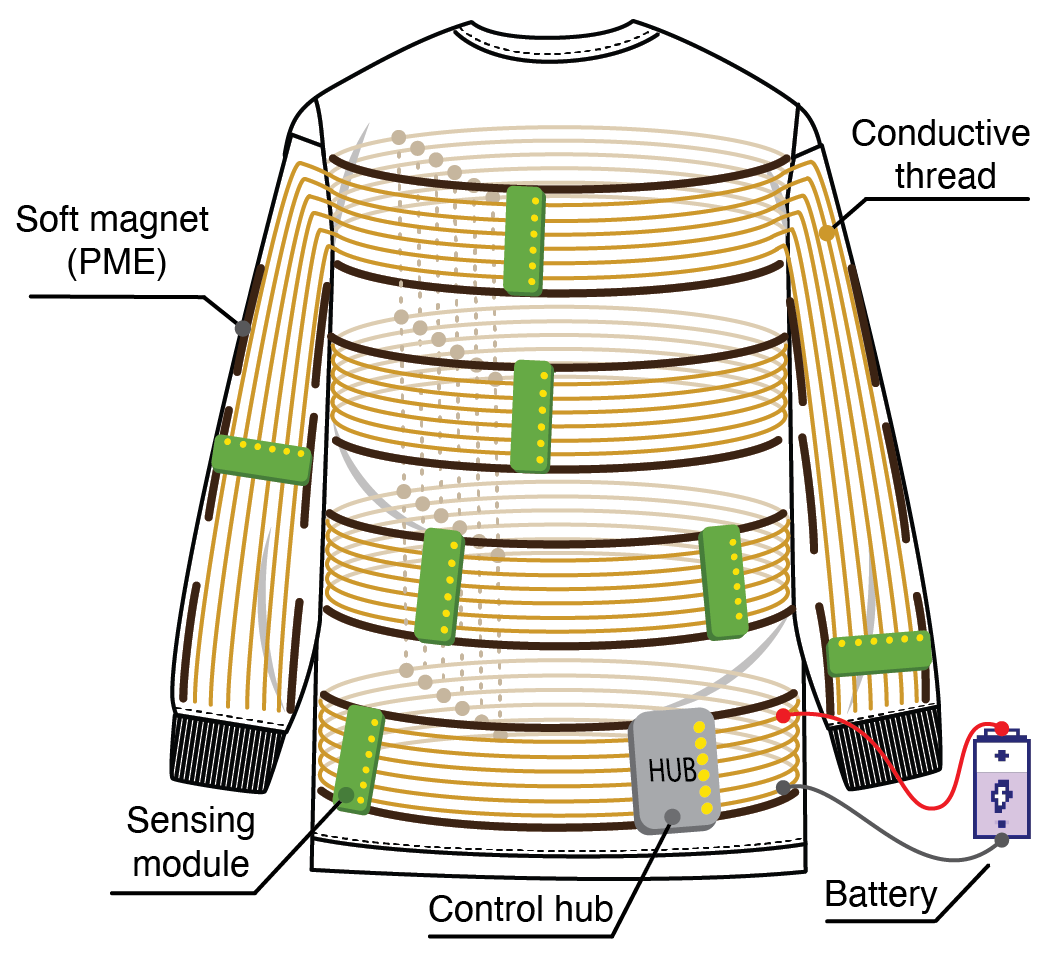}
  \caption{Design overview of plug-n-play e-knit comprising the textile garment, sensing module, soft magnet and control hub.}
  \label{fig:Overview}
  \Description{}
\end{figure}

\begin{figure*}[t!]
  \centering
  \includegraphics[width=1.0\textwidth]{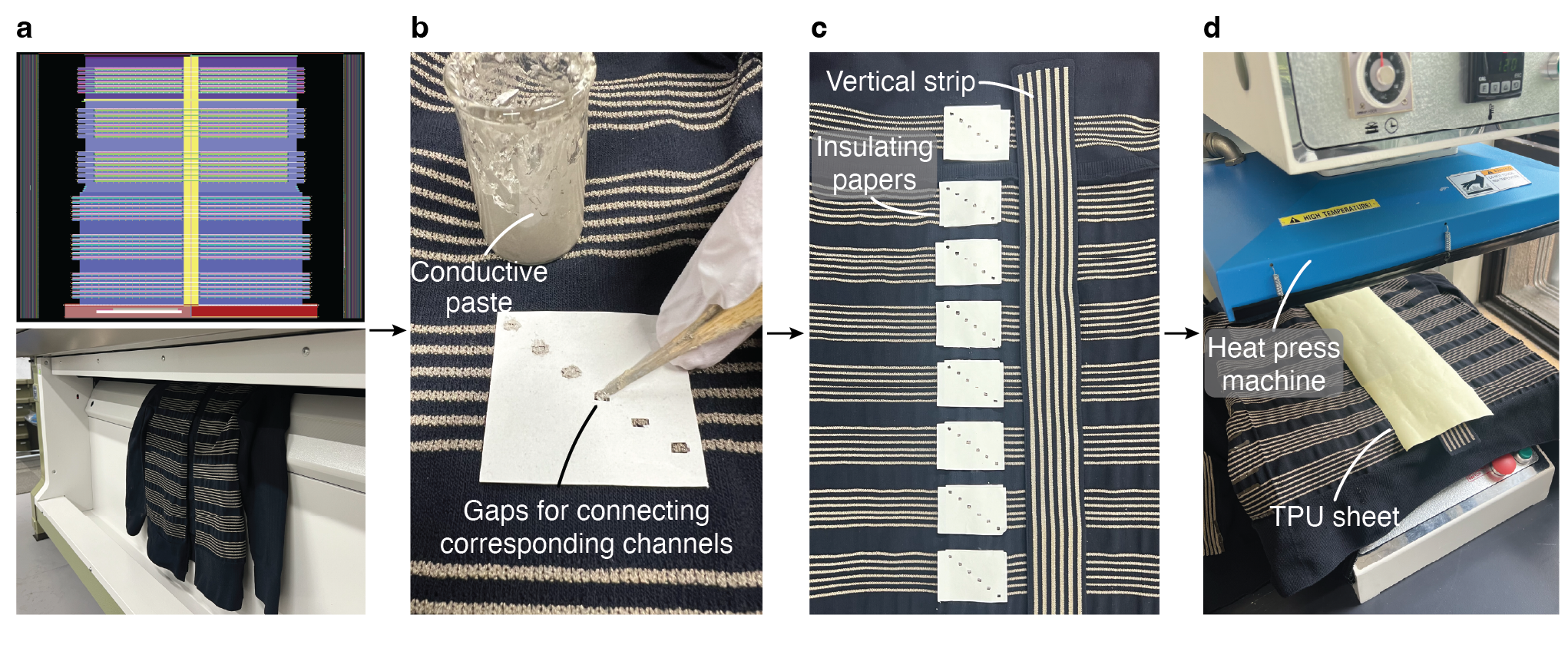}
  \caption{Fabrication process of the textile including (a) designing the thread pattern and producing with industrial knitting machine (MACH2XS, SHIMA SEIKI MFG., LTD.), (b) covering irrelevant channels' conductive threads with square pieces of paper to ensure insulation and coat corresponding channels with conductive paste to enhance conductivity, (c) aligning the gaps in each channel with the corresponding channels of a vertical conductive strip, (d) affixing the strip using Thermoplastic Polyurethanes (TPU) under a heat press machine at \SI{0.7}{\MPa} and \SI{120}{\degreeCelsius}.}
  \label{fig:Textile}
  \Description{}
\end{figure*}

\autoref{fig:Textile}a shows the system overview of plug-n-play e-knit.
Our system consists of three parts: 1) an e-knit with several groups of exposed conductive lines, 2) soft magnet connector which is compatible with a stretchable e-knit, and 3) sensing modules connected with the e-knit via the magnet connectors.

\subsection{Machine-knitted E-textiles}

In \autoref{fig:Overview}, we divide the conductive threads (AGposs, Mitsufuji Corporation)  into several groups, arranged horizontally on the garment and they maintain a certain distance to allow for PME installation and sensing module placement. 
Each group contains six equidistant horizontal conductive threads to form different channels, independent and assigned to transmit different signals. 
The top and bottom channels power the modules, providing a 5V power supply line (i.e., VCC) and a ground line (i.e., GND) signal. 
While The  I$^2$C protocol is promising for plug-n-play e-knit due to its minimal interfaces, fewer pinouts and multiple slave sensors communication,  I$^2$C suffers from noise interference and limited communication distance.
Consequently, we utilized Differential I$^2$C to diminish signal interference on textiles and increase the communication range.
Differential I$^2$C divides the two traditional I$^2$C signals (SDA, SCL) into four differential signals with the same magnitude and opposite polarity (SDA+, SDA-, SCL+, SCL-) for transmission, and integrates them into signals with double amplitude at the receiver end to enhance noise immunity.
These signals are transmitted through 2nd to 5th channels.
Then, we can fabricate the e-textile, as~\autoref{fig:Textile} shows.

\subsection{Soft Magnet Connector}

\begin{figure}[t!]
  \centering
  \includegraphics[width=0.45\textwidth]{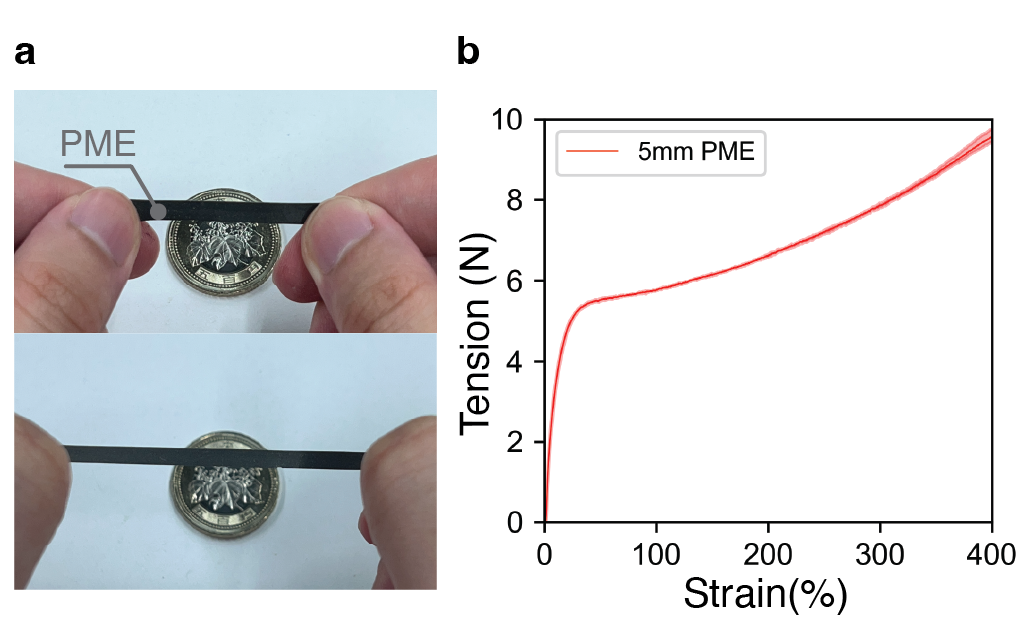}
  \caption{ (a) A close-up of \SI{5}{\mm} wide PME's stretchability. (b)  Relationship between the tension and the strain of the \SI{5}{\mm} wide PME. The solid line denotes the mean and the shaded areas depict the maximum deviation of the three PME testing samples.}
  \label{fig:PME}
  \Description{}
\end{figure}

The plug-and-play e-textile prototyping connector should be lightweight and stretchable, in addition to having a strong connection force.
Prior work has proposed conductive hook-and-loop~\cite{seager2013flexibleconnector}, conductive barb~\cite{woelfle_plug-and-play_2023}, and soft magnets as plug-and-play connectors.
Among them, PME~\cite{Shembekar_PME_2021}, one of the soft magnets, is a stretchable neodymium magnet (see \autoref{fig:PME}a), fabricated by mixing a polyurethane resin with a neodymium powder.
The mixed polyurethane resin is molded into long strips and magnetized.
We use adhesive to attach the PME shaped into long strips at the edge of each wiring group, so that the flexible PCBs equipped with four neodymium magnets can be strongly connected with the wiring. 
We test the stretchability of PME in the Tensile and Compression Testing Machine (MCT-2150W, A\&D Company Ltd.) with three \SI{40}{\mm} $*$ \SI{5}{\mm} sized PME samples. 
The result is shown in~\autoref{fig:PME}b.
The \SI{5}{\mm} PME provides enough stretchability for daily wear.

\begin{figure*}[t!]
  \centering
  \includegraphics[width=1.0\textwidth]{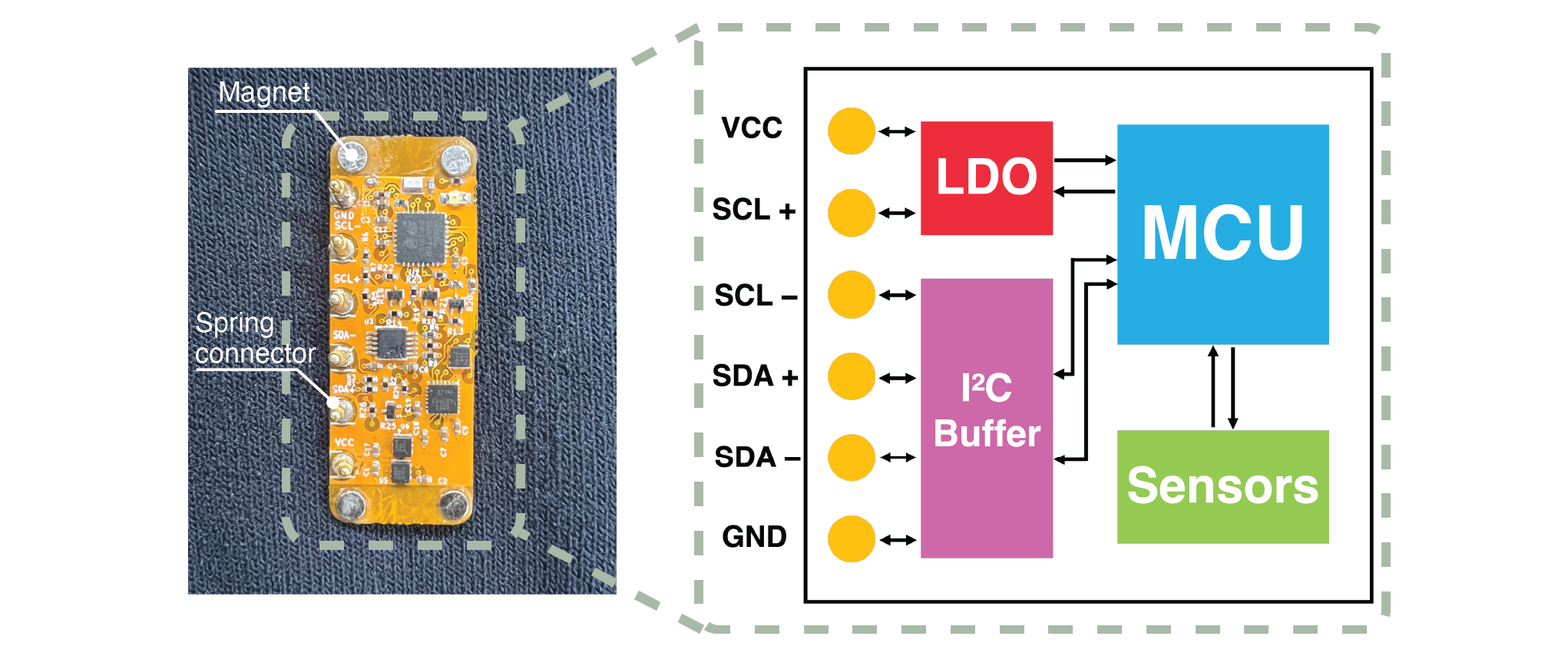}
  \caption{System overview of the sensing module.}
  \label{fig:Sensing}
  \Description{}
\end{figure*}

\subsection{Sensing Module}

To implement various functions on e-textiles while maintaining wear comfort, we need lightweight and customizable sensing modules. 
The energy and signal of the plug-n-play e-knit can be transmitted throughout the entire garment, providing the conditions for this. 
Here, we introduce the IMU sensing module as an example.
As shown in~\autoref{fig:Sensing}, the sensing module consists of flexible PCBs with low-power MCU (STM32L432KC, STMicroelectronics), Low Dropout(LDO) (TLV76733, Texas Instruments), IMU sensor~(ICM20948, TDK InvenSense), Differential I$^2$C buffer chip (PCA9615, NXP Semiconductors) and spring connector which align the position of conductive threads. 
We employ adhesive to attach magnets at the edge of sensing modules for the magnet connection with the PME.
When they are connected, the electric connectors are also in contact with the conductive threads on the textile (\autoref{fig:teaser}c), initiating their operation.
Each module receives power from a central hub and communicates with the hub via the differential I$^2$C protocol. 
The use of lightweight flexible PCB stably records the 9-axis IMU motion data against user motions.
The module is \SI{40}{mm}*\SI{15}{mm} size and \SI{1}{g} weight.
\section{EVALUATION}
This section evaluates the communication and connection stability of the plug-n-play e-knit.

\begin{figure*}[t!]
  \centering
  \includegraphics[width=1.0\textwidth]{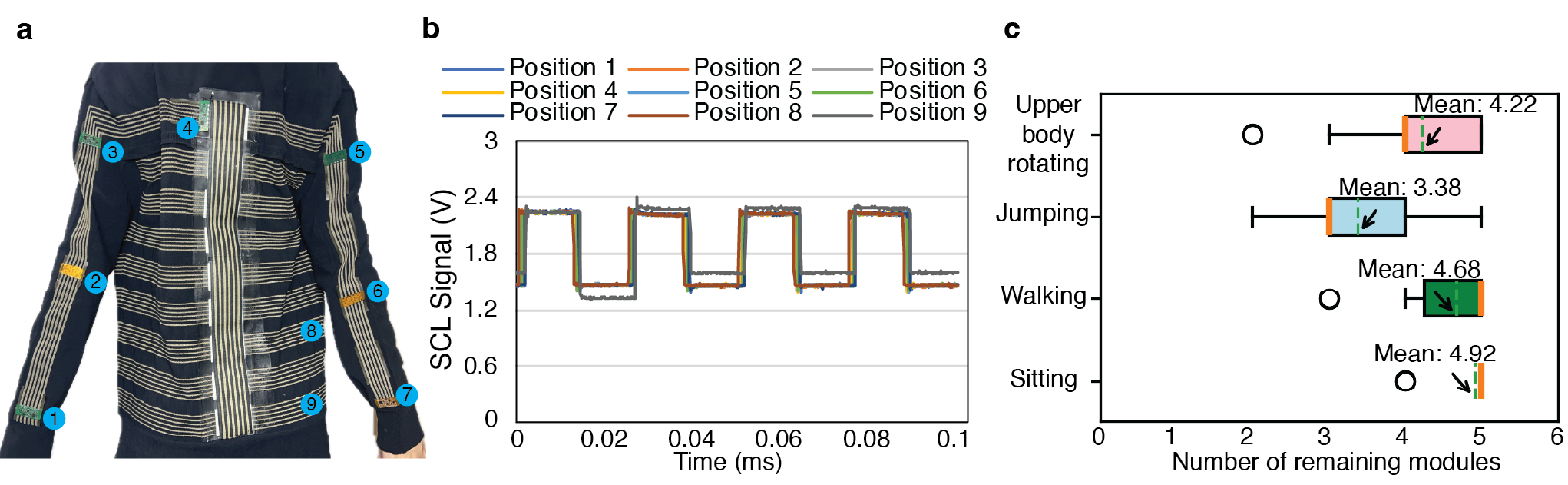}
  \caption{(a) The setup of evaluating 9 position's signal attenuation. (b) Comparison of SCL-signals acquired at 9 locations within~\SI{0.1}{\ms}. (c) 
  Comparison of the number of 5 intact sensing modules left after different movements.}
  \label{fig:Evaluation}
  \Description{}
\end{figure*}

\subsection{Signal Attenuation for Different Body Placement}
\label{sec:eva_signal}

To construct large-scale e-textile prototyping, the maximum communication range between the sensing modules and hubs must be over \SI{1.5}{\m} (i.e., the distance between the left wrist and hem of the garment).
However, the signal attenuation in the long conductive wire causes communication failure.
Therefore, we tested the performance of the two longest channels in the plug-n-play e-knit.
We chose conductive paths between the ends of the two sleeves and one sleeve to the corner of the garment.
We placed one sensing module, which serves as the data transmitter, on the right wrist (Position 1). 
Then, we tested the signal attenuation at the eight positions (Position 2-9), as shown in~\autoref{fig:Evaluation}a.
The voltage of the differential  I$^2$C signal (i.e., SCL signal) at the 8 positions and the transmitting signal from the data transmitter are shown in~\autoref{fig:Evaluation}b. 
Although there is a reduction in the signal at the edge (Position 9), the waveform of the signal does not distort, and the difference between the high and low levels remains clear.
As a result, the plug-n-play e-knit enables the differential  I$^2$C communication without the signal loss.

\subsection{Connection Stability of Soft Magnet Connector}

Since the e-textile typically functions during the user's movement, the soft magnet connection in the plug-n-play e-knit needs to be robust against user motions. 
We evaluated the robustness by monitoring the communication status of the five sensing modules during four types of user motions including walking, running, jumping and upper body rotating.
Note that the sensing modules were placed at the left wrist, right wrist, back, chest, and waist and the motion period was 5 seconds.
\autoref{fig:Evaluation}c shows the remaining number of the sensing modules for each motion test 50 times. 
The result shows that plug-n-play e-knit shows sufficient connection stability except for jumping.
Since the jumping causes the force of \SI{0.03}{\N} to the sensing module, we need to increase the surface magnetic force of the PME by at least 3 times.

\begin{figure*}[!t]
  \centering
  \includegraphics[width=1.0\textwidth]{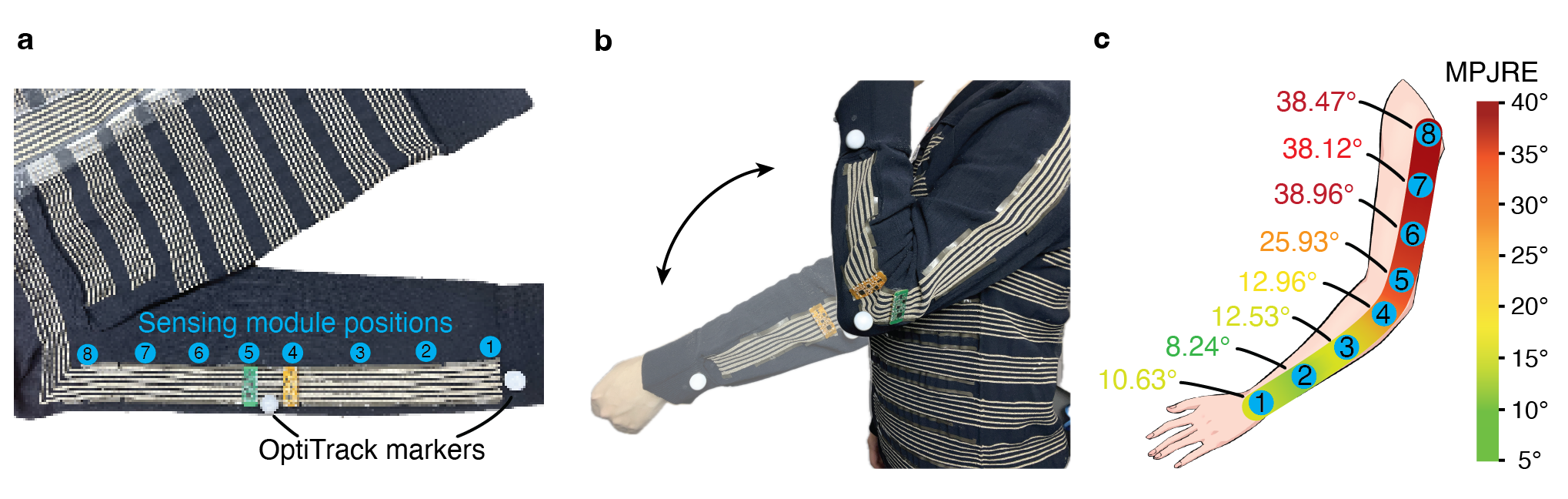}
  \caption{(a) The sensing module is positioned at eight points between the wrist and shoulder to actively test (b) the optimal placement for capturing forearm movement, and (c) get Mean Per Joint Rotation Error (MPJRE) data for each position.}
  \label{fig:Application1}
  \Description{}
\end{figure*}

\section{APPLICATION EXAMPLES}
\subsection{The Optimal Sensor Placement for Motion Capturing}

Previous e-textile prototype tools might necessitate complex circuit connections~\cite{hill_threadboard_2021} or invasive sensor installation methods~\cite{garbacz_modular_2021}, whereas our plug-n-play e-knit offers a large-scale, high-resolution experiment approach.
For example, the user can keep the upper arm stationary while making a forward flexing movement with the forearm. 
Meanwhile, two OptiTrack markers were placed on the wrist and elbow to obtain ground-truth data on forearm movement. 
Then, select eight equidistant positions from the wrist to the shoulder to place the IMU functional module and perform a forearm flexion experiment for each position, as shown in~\autoref{fig:Application1}a, to test which position can get the most accurate forearm movement data.
Since the upper arm remains stationary, we can use the ground truth data as a reference for the angle of forearm flexion and compare the Mean Per Joint Rotation Error (MPJRE) of the IMU at the eight positions.

As shown in~\autoref{fig:Application1}c, we can obtain the most accurate data below the wrist ($8.24^{\circ}$), rather than at the wrist itself ($10.63^{\circ}$). This is due to unconscious wrist twisting during forearm movement. The accuracy of the data diminishes greatly behind the elbow joint. 
During the experiment, the user just needs to plug and unplug the FPCB board to test its performance at different positions on the e-textile. This not only improves the efficiency of prototype design but also reduces the damage to the textile caused by repeated invasive testing. 
Note that the data from above and below the elbow was obtained simultaneously with two IMU modules, indicating that our system supports simultaneous testing with multiple sensing modules. 
Given sufficient modules, we can simultaneously measure the movement at eight positions.

\begin{figure*}[!h]
  \centering
  \includegraphics[width=1.0\textwidth]{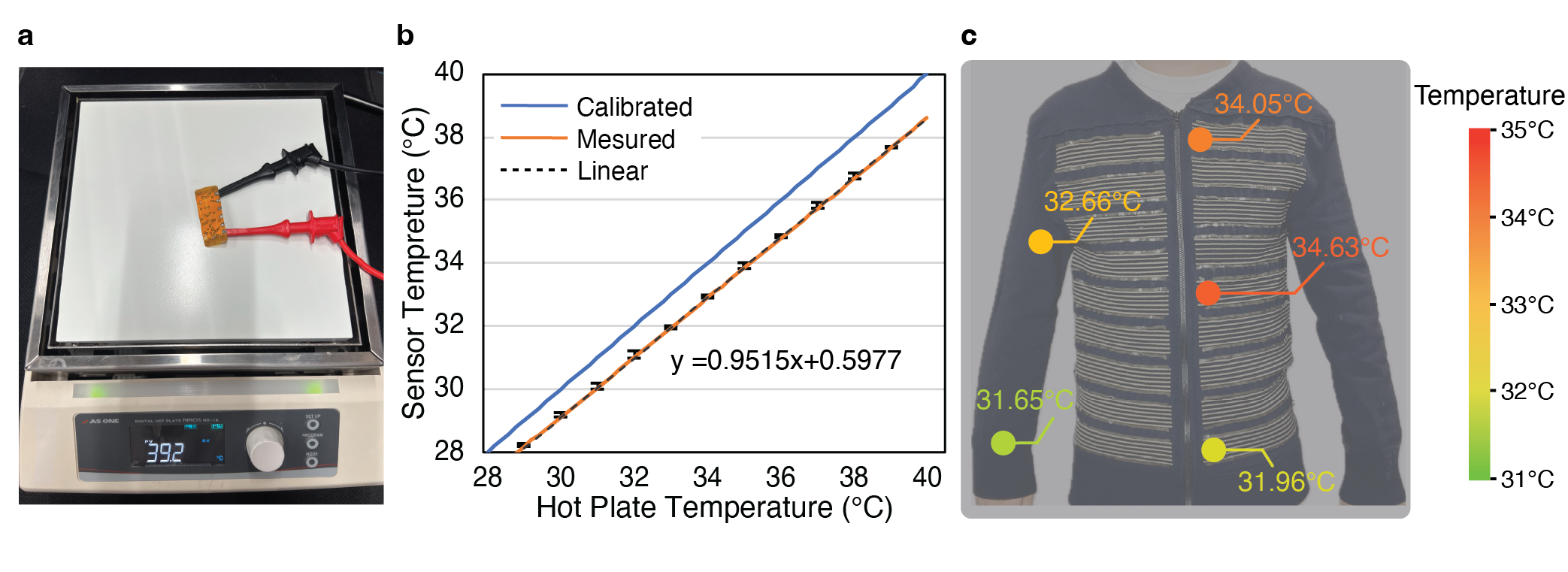}
  \caption{The temperature sensor is (a) placed on a hot plate (ND-1A, AS ONE) and (b) gets the linear fitting expression and graph of temperature characterization with the calibrated result. (c) Then we place the sensing board at 5 positions to get skin temperature data.}
  \label{fig:Application2}
  \Description{}
\end{figure*}

\subsection{Skin Temperature Measurement}
Human skin temperature varies across different body parts~\cite{lee2019regionalVary}. These data can be used to estimate stress levels~\cite{vinkers2013stress}, and understand exercise efficiency~\cite{chudecka2012sportEfficiency}. 
In previous studies, skin temperature was obtained through thermal imaging~\cite{chudecka2012sportEfficiency}, electronic skin~\cite{yokota2015eskin}, and e-textiles~\cite{wicaksono_tailored_2020}. 
These methods either are unsuitable for daily monitoring or do not allow for adjusting the measurement position according to the users' needs. 
The plug-n-play e-knit provides an integrated and reconfigurable on-textile skin temperature measurement method. 
For example, we place the data acquisition hub at the corner of the garment, and then place sensing modules equipped with temperature sensors at the wrist, upper arm, infra-clavicular, and stomach areas and attach them closely to the body for measurement. 
The calibrated results are shown in~\autoref{fig:Application2}c.
During the experiment, users can rearrange the sensing modules in a large area of the garment and get real-time temperature data.

\section{DISCUSSION}
\subsection{Signal Loss against User Movement}

Current e-textiles need the uncovered differential  I$^2$C signal lines for the direct connection of the sensing modules.
Such the exposed signal lines cause the undesired contact of the adjacent signal lines, resulting in the communication error, specifically around the arm.
Using the wireless e-textile only around the arm~\cite{takahashi2021meander, takahashi2022meander} could mitigate this issue by providing the non-contact connection between the wiring-coated e-textiles and the sensing modules.

\subsection{Disconnection of Differential I$^2$C Signal Lines}

As described in Section \ref{sec:eva_signal}, $7.4\%$ area of the differential  I$^2$C signal lines is disconnected with the central hub, owing to the misalignment of the vertical conductive strips with the horizontal Differential  I$^2$C signal lines.
Although we manually adjust the position of the vertical strips when using the heat press, we will explore the stable connection between the vertical and horizontal signal lines with the development of the alignment jig.

\section{CONCLUSION}

This paper introduces the plug-n-play e-knit toward enabling the prototyping tool of the large-area, non-invasive e-textile. 
Combining the machine-knitted e-textile with the soft magnet connector, the plug-n-play e-knit allows users to freely and repeatably rearrange the sensing modules across the e-textiles. 
The plug-n-play e-knit demonstrated its large-area on-textile prototyping ability such as capturing arm movements and measuring skin temperature at scale.
The stable connectivity of the sensing modules against user's harsh movement could be one of our main future work.
We strongly believe the large-scale, non-invasive design of the plug-n-play e-knit could promote the ubiquitous development of smart clothing with various functionalities.

\section{Acknowledgments}

This work was supported by JSPS 22K21343 and JST ACT-X JPMJAX21K9.
The authors thank Resonac Corporation for fabricating PME.

\bibliographystyle{ACM-Reference-Format}
\bibliography{plug-n-play}


\begin{thebibliography}{34}


\ifx \showCODEN    \undefined \def \showCODEN     #1{\unskip}     \fi
\ifx \showDOI      \undefined \def \showDOI       #1{#1}\fi
\ifx \showISBNx    \undefined \def \showISBNx     #1{\unskip}     \fi
\ifx \showISBNxiii \undefined \def \showISBNxiii  #1{\unskip}     \fi
\ifx \showISSN     \undefined \def \showISSN      #1{\unskip}     \fi
\ifx \showLCCN     \undefined \def \showLCCN      #1{\unskip}     \fi
\ifx \shownote     \undefined \def \shownote      #1{#1}          \fi
\ifx \showarticletitle \undefined \def \showarticletitle #1{#1}   \fi
\ifx \showURL      \undefined \def \showURL       {\relax}        \fi
\providecommand\bibfield[2]{#2}
\providecommand\bibinfo[2]{#2}
\providecommand\natexlab[1]{#1}
\providecommand\showeprint[2][]{arXiv:#2}

\bibitem[Buechley and Eisenberg(2008)]%
        {buechley_lilypad_2008}
\bibfield{author}{\bibinfo{person}{Leah Buechley} {and} \bibinfo{person}{Michael Eisenberg}.} \bibinfo{year}{2008}\natexlab{}.
\newblock \showarticletitle{The {LilyPad} {Arduino}: {Toward} {Wearable} {Engineering} for {Everyone}}.
\newblock \bibinfo{journal}{\emph{IEEE Pervasive Computing}} \bibinfo{volume}{7}, \bibinfo{number}{2} (\bibinfo{date}{April} \bibinfo{year}{2008}), \bibinfo{pages}{12--15}.
\newblock
\showISSN{1558-2590}
\urldef\tempurl%
\url{https://doi.org/10.1109/MPRV.2008.38}
\showDOI{\tempurl}


\bibitem[Chudecka and Lubkowska(2012)]%
        {chudecka2012sportEfficiency}
\bibfield{author}{\bibinfo{person}{Monika Chudecka} {and} \bibinfo{person}{Anna Lubkowska}.} \bibinfo{year}{2012}\natexlab{}.
\newblock \showarticletitle{The use of thermal imaging to evaluate body temperature changes of athletes during training and a study on the impact of physiological and morphological factors on skin temperature}.
\newblock \bibinfo{journal}{\emph{Human movement}} \bibinfo{volume}{13}, \bibinfo{number}{1} (\bibinfo{year}{2012}), \bibinfo{pages}{33--39}.
\newblock


\bibitem[Fobelets et~al\mbox{.}(2019)]%
        {fobelets2019wirelessknit}
\bibfield{author}{\bibinfo{person}{Kristel Fobelets}, \bibinfo{person}{Kris Thielemans}, \bibinfo{person}{Abhinaya Mathivanan}, {and} \bibinfo{person}{Christos Papavassiliou}.} \bibinfo{year}{2019}\natexlab{}.
\newblock \showarticletitle{Characterization of Knitted Coils for e-Textiles}.
\newblock \bibinfo{journal}{\emph{IEEE Sensors Journal}} \bibinfo{volume}{19}, \bibinfo{number}{18} (\bibinfo{year}{2019}), \bibinfo{pages}{7835--7840}.
\newblock
\urldef\tempurl%
\url{https://doi.org/10.1109/JSEN.2019.2917542}
\showDOI{\tempurl}


\bibitem[Fromme et~al\mbox{.}(2021)]%
        {fromme_metal-textile_2021}
\bibfield{author}{\bibinfo{person}{Nicolas~P. Fromme}, \bibinfo{person}{Yifan Li}, \bibinfo{person}{Martin Camenzind}, \bibinfo{person}{Claudio Toncelli}, {and} \bibinfo{person}{René~M. Rossi}.} \bibinfo{year}{2021}\natexlab{}.
\newblock \showarticletitle{Metal-{Textile} {Laser} {Welding} for {Wearable} {Sensors} {Applications}}.
\newblock \bibinfo{journal}{\emph{Advanced Electronic Materials}} \bibinfo{volume}{7}, \bibinfo{number}{4} (\bibinfo{year}{2021}), \bibinfo{pages}{2001238}.
\newblock
\showISSN{2199-160X}
\urldef\tempurl%
\url{https://doi.org/10.1002/aelm.202001238}
\showDOI{\tempurl}


\bibitem[Garbacz et~al\mbox{.}(2021)]%
        {garbacz_modular_2021}
\bibfield{author}{\bibinfo{person}{Kamil Garbacz}, \bibinfo{person}{Lars Stagun}, \bibinfo{person}{Sigrid Rotzler}, \bibinfo{person}{Markus Semenec}, {and} \bibinfo{person}{Malte von Krshiwoblozki}.} \bibinfo{year}{2021}\natexlab{}.
\newblock \showarticletitle{Modular {E}-{Textile} {Toolkit} for {Prototyping} and {Manufacturing}}.
\newblock \bibinfo{journal}{\emph{Proceedings}} \bibinfo{volume}{68}, \bibinfo{number}{1} (\bibinfo{year}{2021}), \bibinfo{pages}{5}.
\newblock
\showISSN{2504-3900}
\urldef\tempurl%
\url{https://doi.org/10.3390/proceedings2021068005}
\showDOI{\tempurl}


\bibitem[Griffin et~al\mbox{.}(2016)]%
        {griffin_analysis_2016}
\bibfield{author}{\bibinfo{person}{Linsey Griffin}, \bibinfo{person}{Crystal Compton}, {and} \bibinfo{person}{Lucy~E. Dunne}.} \bibinfo{year}{2016}\natexlab{}.
\newblock \showarticletitle{An analysis of the variability of anatomical body references within ready-to-wear garment sizes}. In \bibinfo{booktitle}{\emph{Proceedings of the 2016 {ACM} {International} {Symposium} on {Wearable} {Computers}}} \emph{(\bibinfo{series}{{ISWC} '16})}. \bibinfo{publisher}{Association for Computing Machinery}, \bibinfo{address}{New York, NY, USA}, \bibinfo{pages}{84--91}.
\newblock
\showISBNx{978-1-4503-4460-9}
\urldef\tempurl%
\url{https://doi.org/10.1145/2971763.2971800}
\showDOI{\tempurl}


\bibitem[Hamdan et~al\mbox{.}(2018)]%
        {hamdan2018embridery}
\bibfield{author}{\bibinfo{person}{Nur Al-huda Hamdan}, \bibinfo{person}{Simon Voelker}, {and} \bibinfo{person}{Jan Borchers}.} \bibinfo{year}{2018}\natexlab{}.
\newblock \showarticletitle{Sketch Stitch: Interactive Embroidery for E-textiles}. In \bibinfo{booktitle}{\emph{Proceedings of the 2018 CHI Conference on Human Factors in Computing Systems}} (Montreal QC, Canada) \emph{(\bibinfo{series}{CHI '18})}. \bibinfo{publisher}{Association for Computing Machinery}, \bibinfo{address}{New York, NY, USA}, \bibinfo{pages}{1–13}.
\newblock
\showISBNx{9781450356206}
\urldef\tempurl%
\url{https://doi.org/10.1145/3173574.3173656}
\showDOI{\tempurl}


\bibitem[Han et~al\mbox{.}(2021)]%
        {han_towards_2021}
\bibfield{author}{\bibinfo{person}{Ahyoung Han}, \bibinfo{person}{Kwangyun Wohn}, {and} \bibinfo{person}{Jaehong Ahn}.} \bibinfo{year}{2021}\natexlab{}.
\newblock \showarticletitle{Towards new fashion design education: learning virtual prototyping using {E}-textiles}.
\newblock \bibinfo{journal}{\emph{International Journal of Technology and Design Education}} \bibinfo{volume}{31}, \bibinfo{number}{2} (\bibinfo{date}{April} \bibinfo{year}{2021}), \bibinfo{pages}{379--400}.
\newblock
\showISSN{1573-1804}
\urldef\tempurl%
\url{https://doi.org/10.1007/s10798-019-09558-w}
\showDOI{\tempurl}


\bibitem[Hill et~al\mbox{.}(2021)]%
        {hill_threadboard_2021}
\bibfield{author}{\bibinfo{person}{Chris Hill}, \bibinfo{person}{Michael Schneider}, \bibinfo{person}{Ann Eisenberg}, {and} \bibinfo{person}{Mark~D. Gross}.} \bibinfo{year}{2021}\natexlab{}.
\newblock \showarticletitle{The {ThreadBoard}: {Designing} an {E}-{Textile} {Rapid} {Prototyping} {Board}}. In \bibinfo{booktitle}{\emph{Proceedings of the {Fifteenth} {International} {Conference} on {Tangible}, {Embedded}, and {Embodied} {Interaction}}} \emph{(\bibinfo{series}{{TEI} '21})}. \bibinfo{publisher}{Association for Computing Machinery}, \bibinfo{address}{New York, NY, USA}, \bibinfo{pages}{1--7}.
\newblock
\showISBNx{978-1-4503-8213-7}
\urldef\tempurl%
\url{https://doi.org/10.1145/3430524.3440642}
\showDOI{\tempurl}


\bibitem[Jones et~al\mbox{.}(2020a)]%
        {jones_swatch-bits_2020}
\bibfield{author}{\bibinfo{person}{Lee Jones}, \bibinfo{person}{Sara Nabil}, {and} \bibinfo{person}{Audrey Girouard}.} \bibinfo{year}{2020}\natexlab{a}.
\newblock \showarticletitle{Swatch-bits: {Prototyping} {E}-textiles with {Modular} {Swatches}}. In \bibinfo{booktitle}{\emph{Proceedings of the {Fourteenth} {International} {Conference} on {Tangible}, {Embedded}, and {Embodied} {Interaction}}} \emph{(\bibinfo{series}{{TEI} '20})}. \bibinfo{publisher}{Association for Computing Machinery}, \bibinfo{address}{New York, NY, USA}, \bibinfo{pages}{893--897}.
\newblock
\showISBNx{978-1-4503-6107-1}
\urldef\tempurl%
\url{https://doi.org/10.1145/3374920.3374971}
\showDOI{\tempurl}


\bibitem[Jones et~al\mbox{.}(2020b)]%
        {jones_wearable_2020}
\bibfield{author}{\bibinfo{person}{Lee Jones}, \bibinfo{person}{Sara Nabil}, \bibinfo{person}{Amanda McLeod}, {and} \bibinfo{person}{Audrey Girouard}.} \bibinfo{year}{2020}\natexlab{b}.
\newblock \showarticletitle{Wearable {Bits}: {Scaffolding} {Creativity} with a {Prototyping} {Toolkit} for {Wearable} {E}-textiles}. In \bibinfo{booktitle}{\emph{Proceedings of the {Fourteenth} {International} {Conference} on {Tangible}, {Embedded}, and {Embodied} {Interaction}}} \emph{(\bibinfo{series}{{TEI} '20})}. \bibinfo{publisher}{Association for Computing Machinery}, \bibinfo{address}{New York, NY, USA}, \bibinfo{pages}{165--177}.
\newblock
\showISBNx{978-1-4503-6107-1}
\urldef\tempurl%
\url{https://doi.org/10.1145/3374920.3374954}
\showDOI{\tempurl}


\bibitem[Lee et~al\mbox{.}(2019)]%
        {lee2019regionalVary}
\bibfield{author}{\bibinfo{person}{Chan~Mi Lee}, \bibinfo{person}{Seon-Pil Jin}, \bibinfo{person}{Eun~Jin Doh}, \bibinfo{person}{Dong~Hun Lee}, {and} \bibinfo{person}{Jin~Ho Chung}.} \bibinfo{year}{2019}\natexlab{}.
\newblock \showarticletitle{Regional variation of human skin surface temperature}.
\newblock \bibinfo{journal}{\emph{Annals of Dermatology}} \bibinfo{volume}{31}, \bibinfo{number}{3} (\bibinfo{year}{2019}), \bibinfo{pages}{349--352}.
\newblock


\bibitem[Leśnikowski(2016)]%
        {lesnikowski_research_2016}
\bibfield{author}{\bibinfo{person}{Jacek Leśnikowski}.} \bibinfo{year}{2016}\natexlab{}.
\newblock \showarticletitle{Research on {Poppers} {Used} as {Electrical} {Connectors} in {High} {Speed} {Textile} {Transmission} {Lines}}.
\newblock \bibinfo{journal}{\emph{Autex Research Journal}} \bibinfo{volume}{16}, \bibinfo{number}{4} (\bibinfo{year}{2016}), \bibinfo{pages}{228--235}.
\newblock
\urldef\tempurl%
\url{https://doi.org/doi:10.1515/aut-2016-0025}
\showDOI{\tempurl}


\bibitem[Li et~al\mbox{.}(2018)]%
        {li2018printcoil}
\bibfield{author}{\bibinfo{person}{Yi Li}, \bibinfo{person}{Neil Grabham}, \bibinfo{person}{Russel Torah}, \bibinfo{person}{John Tudor}, {and} \bibinfo{person}{Steve Beeby}.} \bibinfo{year}{2018}\natexlab{}.
\newblock \showarticletitle{Textile-based flexible coils for wireless inductive power transmission}.
\newblock \bibinfo{journal}{\emph{Applied Sciences}} \bibinfo{volume}{8}, \bibinfo{number}{6} (\bibinfo{year}{2018}), \bibinfo{pages}{912}.
\newblock


\bibitem[Lin et~al\mbox{.}(2022)]%
        {lin_digitally-embroidered_2022}
\bibfield{author}{\bibinfo{person}{Rongzhou Lin}, \bibinfo{person}{Han-Joon Kim}, \bibinfo{person}{Sippanat Achavananthadith}, \bibinfo{person}{Ze Xiong}, \bibinfo{person}{Jason K.~W. Lee}, \bibinfo{person}{Yong~Lin Kong}, {and} \bibinfo{person}{John~S. Ho}.} \bibinfo{year}{2022}\natexlab{}.
\newblock \showarticletitle{Digitally-embroidered liquid metal electronic textiles for wearable wireless systems}.
\newblock \bibinfo{journal}{\emph{Nature Communications}} \bibinfo{volume}{13}, \bibinfo{number}{1} (\bibinfo{date}{April} \bibinfo{year}{2022}), \bibinfo{pages}{2190}.
\newblock
\showISSN{2041-1723}
\urldef\tempurl%
\url{https://doi.org/10.1038/s41467-022-29859-4}
\showDOI{\tempurl}


\bibitem[Linz et~al\mbox{.}(2005)]%
        {linz2005stitching}
\bibfield{author}{\bibinfo{person}{Torsten Linz}, \bibinfo{person}{Christine Kallmayer}, \bibinfo{person}{Rolf Aschenbrenner}, {and} \bibinfo{person}{Herbert Reichl}.} \bibinfo{year}{2005}\natexlab{}.
\newblock \showarticletitle{Embroidering electrical interconnects with conductive yarn for the integration of flexible electronic modules into fabric}. In \bibinfo{booktitle}{\emph{Ninth IEEE International Symposium on Wearable Computers (ISWC'05)}}. IEEE, \bibinfo{pages}{86--89}.
\newblock


\bibitem[Markvicka et~al\mbox{.}(2018)]%
        {markvicka_STEAM_2018}
\bibfield{author}{\bibinfo{person}{Eric Markvicka}, \bibinfo{person}{Steven Rich}, \bibinfo{person}{Jiahe Liao}, \bibinfo{person}{Hesham Zaini}, {and} \bibinfo{person}{Carmel Majidi}.} \bibinfo{year}{2018}\natexlab{}.
\newblock \showarticletitle{Low-cost wearable human-computer interface with conductive fabric for STEAM education}. In \bibinfo{booktitle}{\emph{2018 IEEE Integrated STEM Education Conference (ISEC)}}. \bibinfo{pages}{161--166}.
\newblock
\urldef\tempurl%
\url{https://doi.org/10.1109/ISECon.2018.8340469}
\showDOI{\tempurl}


\bibitem[Noda and Shinoda(2019)]%
        {noda_I2We_2019}
\bibfield{author}{\bibinfo{person}{Akihito Noda} {and} \bibinfo{person}{Hiroyuki Shinoda}.} \bibinfo{year}{2019}\natexlab{}.
\newblock \showarticletitle{Inter-IC for Wearables (I2We): Power and Data Transfer Over Double-Sided Conductive Textile}.
\newblock \bibinfo{journal}{\emph{IEEE Transactions on Biomedical Circuits and Systems}} \bibinfo{volume}{13}, \bibinfo{number}{1} (\bibinfo{year}{2019}), \bibinfo{pages}{80--90}.
\newblock
\urldef\tempurl%
\url{https://doi.org/10.1109/TBCAS.2018.2881219}
\showDOI{\tempurl}


\bibitem[Ohiri et~al\mbox{.}(2022)]%
        {ohiri_e-textile_2022}
\bibfield{author}{\bibinfo{person}{Korine~A. Ohiri}, \bibinfo{person}{Connor~O. Pyles}, \bibinfo{person}{Leslie~H. Hamilton}, \bibinfo{person}{Megan~M. Baker}, \bibinfo{person}{Matthew~T. McGuire}, \bibinfo{person}{Eric~Q. Nguyen}, \bibinfo{person}{Luke~E. Osborn}, \bibinfo{person}{Katelyn~M. Rossick}, \bibinfo{person}{Emil~G. McDowell}, \bibinfo{person}{Leah~M. Strohsnitter}, {and} \bibinfo{person}{Luke~J. Currano}.} \bibinfo{year}{2022}\natexlab{}.
\newblock \showarticletitle{E-textile based modular {sEMG} suit for large area level of effort analysis}.
\newblock \bibinfo{journal}{\emph{Scientific Reports}} \bibinfo{volume}{12}, \bibinfo{number}{1} (\bibinfo{date}{June} \bibinfo{year}{2022}), \bibinfo{pages}{9650}.
\newblock
\showISSN{2045-2322}
\urldef\tempurl%
\url{https://doi.org/10.1038/s41598-022-13701-4}
\showDOI{\tempurl}


\bibitem[Seager et~al\mbox{.}(2013)]%
        {seager2013flexibleconnector}
\bibfield{author}{\bibinfo{person}{RD Seager}, \bibinfo{person}{Alford Chauraya}, \bibinfo{person}{Shiyu Zhang}, \bibinfo{person}{William Whittow}, {and} \bibinfo{person}{Y Vardaxoglou}.} \bibinfo{year}{2013}\natexlab{}.
\newblock \showarticletitle{Flexible radio frequency connectors for textile electronics}.
\newblock \bibinfo{journal}{\emph{Electronics Letters}} \bibinfo{volume}{49}, \bibinfo{number}{22} (\bibinfo{year}{2013}), \bibinfo{pages}{1371--1373}.
\newblock


\bibitem[Sensoria Fitness. n.d.({[n.\,d.]})]%
        {socks}
Sensoria Fitness. n.d. \bibinfo{year}{[n.\,d.]}\natexlab{}.
\newblock
\newblock
\urldef\tempurl%
\url{https://www.sensoriafitness.com/}
\showURL{%
\tempurl}


\bibitem[Shembekar et~al\mbox{.}(2021)]%
        {Shembekar_PME_2021}
\bibfield{author}{\bibinfo{person}{Sahil Shembekar}, \bibinfo{person}{Mitsuhiro Kamezaki}, \bibinfo{person}{Peizhi Zhang}, \bibinfo{person}{He Zhuoyi}, \bibinfo{person}{Yuhiro Iwamoto}, \bibinfo{person}{Yasushi Ido}, \bibinfo{person}{Hiroyuki Sakamoto}, {and} \bibinfo{person}{Shigeki Sugano}.} \bibinfo{year}{2021}\natexlab{}.
\newblock \showarticletitle{Development of a Permanent Magnet Elastomer (PME) Infused Soft Robot Skin for Tactile Sensing}. \bibinfo{pages}{6039--6046}.
\newblock
\urldef\tempurl%
\url{https://doi.org/10.1109/IROS51168.2021.9636817}
\showDOI{\tempurl}


\bibitem[Song et~al\mbox{.}(2010)]%
        {song2010knit}
\bibfield{author}{\bibinfo{person}{H-Y Song}, \bibinfo{person}{J-H Lee}, \bibinfo{person}{D Kang}, \bibinfo{person}{H Cho}, \bibinfo{person}{H-S Cho}, \bibinfo{person}{J-W Lee}, {and} \bibinfo{person}{Y-J Lee}.} \bibinfo{year}{2010}\natexlab{}.
\newblock \showarticletitle{Textile electrodes of jacquard woven fabrics for biosignal measurement}.
\newblock \bibinfo{journal}{\emph{The Journal of the Textile Institute}} \bibinfo{volume}{101}, \bibinfo{number}{8} (\bibinfo{year}{2010}), \bibinfo{pages}{758--770}.
\newblock


\bibitem[Stanley et~al\mbox{.}(2022a)]%
        {stanley_modular_2022}
\bibfield{author}{\bibinfo{person}{Jessica Stanley}, \bibinfo{person}{Katy Griggs}, \bibinfo{person}{Oliver Handford}, \bibinfo{person}{John~A. Hunt}, \bibinfo{person}{Phil Kunovski}, {and} \bibinfo{person}{Yang Wei}.} \bibinfo{year}{2022}\natexlab{a}.
\newblock \showarticletitle{Modular {E}-{Textile} {Platform} for {Real}-{Time} {Sensing}}. In \bibinfo{booktitle}{\emph{Proceedings of the 2022 {ACM} {International} {Symposium} on {Wearable} {Computers}}} \emph{(\bibinfo{series}{{ISWC} '22})}. \bibinfo{publisher}{Association for Computing Machinery}, \bibinfo{address}{New York, NY, USA}, \bibinfo{pages}{131--135}.
\newblock
\showISBNx{978-1-4503-9424-6}
\urldef\tempurl%
\url{https://doi.org/10.1145/3544794.3560293}
\showDOI{\tempurl}


\bibitem[Stanley et~al\mbox{.}(2022b)]%
        {stanley_review_2022}
\bibfield{author}{\bibinfo{person}{Jessica Stanley}, \bibinfo{person}{John~A. Hunt}, \bibinfo{person}{Phil Kunovski}, {and} \bibinfo{person}{Yang Wei}.} \bibinfo{year}{2022}\natexlab{b}.
\newblock \showarticletitle{A review of connectors and joining technologies for electronic textiles}.
\newblock \bibinfo{journal}{\emph{Engineering Reports}} \bibinfo{volume}{4}, \bibinfo{number}{6} (\bibinfo{year}{2022}), \bibinfo{pages}{e12491}.
\newblock
\showISSN{2577-8196}
\urldef\tempurl%
\url{https://doi.org/10.1002/eng2.12491}
\showDOI{\tempurl}


\bibitem[Takahashi et~al\mbox{.}(2018)]%
        {takahashi2018cuttable}
\bibfield{author}{\bibinfo{person}{Ryo Takahashi}, \bibinfo{person}{Takuya Sasatani}, \bibinfo{person}{Fuminori Okuya}, \bibinfo{person}{Yoshiaki Narusue}, {and} \bibinfo{person}{Yoshihiro Kawahara}.} \bibinfo{year}{2018}\natexlab{}.
\newblock \showarticletitle{A Cuttable Wireless Power Transfer Sheet}.
\newblock \bibinfo{journal}{\emph{Proc. ACM Interact. Mob. Wearable Ubiquitous Technol.}} \bibinfo{volume}{2}, \bibinfo{number}{4}, Article \bibinfo{articleno}{190} (\bibinfo{date}{Dec.} \bibinfo{year}{2018}), \bibinfo{numpages}{25}~pages.
\newblock
\urldef\tempurl%
\url{https://doi.org/10.1145/3287068}
\showDOI{\tempurl}


\bibitem[Takahashi et~al\mbox{.}(2022a)]%
        {takahashi2021meander}
\bibfield{author}{\bibinfo{person}{Ryo Takahashi}, \bibinfo{person}{Wakako Yukita}, \bibinfo{person}{Takuya Sasatani}, \bibinfo{person}{Tomoyuki Yokota}, \bibinfo{person}{Takao Someya}, {and} \bibinfo{person}{Yoshihiro Kawahara}.} \bibinfo{year}{2022}\natexlab{a}.
\newblock \showarticletitle{Twin Meander Coil: Sensitive Readout of Battery-free On-body Wireless Sensors Using Body-scale Meander Coils}.
\newblock \bibinfo{journal}{\emph{Proc. ACM Interact. Mob. Wearable Ubiquitous Technol.}} \bibinfo{volume}{5}, \bibinfo{number}{4}, Article \bibinfo{articleno}{179} (\bibinfo{date}{Dec.} \bibinfo{year}{2022}), \bibinfo{numpages}{21}~pages.
\newblock
\urldef\tempurl%
\url{https://doi.org/10.1145/3494996}
\showDOI{\tempurl}


\bibitem[Takahashi et~al\mbox{.}(2022b)]%
        {takahashi2022meander}
\bibfield{author}{\bibinfo{person}{Ryo Takahashi}, \bibinfo{person}{Wakako Yukita}, \bibinfo{person}{Tomoyuki Yokota}, \bibinfo{person}{Takao Someya}, {and} \bibinfo{person}{Yoshihiro Kawahara}.} \bibinfo{year}{2022}\natexlab{b}.
\newblock \showarticletitle{Meander Coil++: A Body-scale Wireless Power Transmission Using Safe-to-body and Energy-efficient Transmitter Coil}. In \bibinfo{booktitle}{\emph{Proceedings of the 2022 CHI Conference on Human Factors in Computing Systems}} (New Orleans, LA, USA) \emph{(\bibinfo{series}{CHI '22})}. \bibinfo{publisher}{Association for Computing Machinery}, \bibinfo{address}{New York, NY, USA}, Article \bibinfo{articleno}{390}, \bibinfo{numpages}{12}~pages.
\newblock
\showISBNx{9781450391573}
\urldef\tempurl%
\url{https://doi.org/10.1145/3491102.3502119}
\showDOI{\tempurl}


\bibitem[Tseghai et~al\mbox{.}(2020)]%
        {tseghai2020review}
\bibfield{author}{\bibinfo{person}{Granch~Berhe Tseghai}, \bibinfo{person}{Benny Malengier}, \bibinfo{person}{Kinde~Anlay Fante}, \bibinfo{person}{Abreha~Bayrau Nigusse}, {and} \bibinfo{person}{Lieva Van~Langenhove}.} \bibinfo{year}{2020}\natexlab{}.
\newblock \showarticletitle{Integration of conductive materials with textile structures, an overview}.
\newblock \bibinfo{journal}{\emph{Sensors}} \bibinfo{volume}{20}, \bibinfo{number}{23} (\bibinfo{year}{2020}), \bibinfo{pages}{6910}.
\newblock


\bibitem[Vinkers et~al\mbox{.}(2013)]%
        {vinkers2013stress}
\bibfield{author}{\bibinfo{person}{Christiaan~H Vinkers}, \bibinfo{person}{Renske Penning}, \bibinfo{person}{Juliane Hellhammer}, \bibinfo{person}{Joris~C Verster}, \bibinfo{person}{John~HGM Klaessens}, \bibinfo{person}{Berend Olivier}, {and} \bibinfo{person}{Cor~J Kalkman}.} \bibinfo{year}{2013}\natexlab{}.
\newblock \showarticletitle{The effect of stress on core and peripheral body temperature in humans}.
\newblock \bibinfo{journal}{\emph{Stress}} \bibinfo{volume}{16}, \bibinfo{number}{5} (\bibinfo{year}{2013}), \bibinfo{pages}{520--530}.
\newblock


\bibitem[Wang et~al\mbox{.}(2023)]%
        {Wang2023Emtex}
\bibfield{author}{\bibinfo{person}{Qi Wang}, \bibinfo{person}{Yuan Zeng}, \bibinfo{person}{Runhua Zhang}, \bibinfo{person}{Nianding Ye}, \bibinfo{person}{Linghao Zhu}, \bibinfo{person}{Xiaohua Sun}, {and} \bibinfo{person}{Teng Han}.} \bibinfo{year}{2023}\natexlab{}.
\newblock \showarticletitle{EmTex: Prototyping Textile-Based Interfaces through An Embroidered Construction Kit}. In \bibinfo{booktitle}{\emph{Proceedings of the 36th Annual ACM Symposium on User Interface Software and Technology}} (San Francisco, CA, USA) \emph{(\bibinfo{series}{UIST '23})}. \bibinfo{publisher}{Association for Computing Machinery}, \bibinfo{address}{New York, NY, USA}, Article \bibinfo{articleno}{20}, \bibinfo{numpages}{17}~pages.
\newblock
\showISBNx{9798400701320}
\urldef\tempurl%
\url{https://doi.org/10.1145/3586183.3606815}
\showDOI{\tempurl}


\bibitem[Wicaksono et~al\mbox{.}(2020)]%
        {wicaksono_tailored_2020}
\bibfield{author}{\bibinfo{person}{Irmandy Wicaksono}, \bibinfo{person}{Carson~I. Tucker}, \bibinfo{person}{Tao Sun}, \bibinfo{person}{Cesar~A. Guerrero}, \bibinfo{person}{Clare Liu}, \bibinfo{person}{Wesley~M. Woo}, \bibinfo{person}{Eric~J. Pence}, {and} \bibinfo{person}{Canan Dagdeviren}.} \bibinfo{year}{2020}\natexlab{}.
\newblock \showarticletitle{A tailored, electronic textile conformable suit for large-scale spatiotemporal physiological sensing in vivo}.
\newblock \bibinfo{journal}{\emph{npj Flexible Electronics}} \bibinfo{volume}{4}, \bibinfo{number}{1} (\bibinfo{date}{April} \bibinfo{year}{2020}), \bibinfo{pages}{1--13}.
\newblock
\showISSN{2397-4621}
\urldef\tempurl%
\url{https://doi.org/10.1038/s41528-020-0068-y}
\showDOI{\tempurl}


\bibitem[Woelfle et~al\mbox{.}(2023)]%
        {woelfle_plug-and-play_2023}
\bibfield{author}{\bibinfo{person}{Heidi Woelfle}, \bibinfo{person}{Olaitan Adeleke}, \bibinfo{person}{Niharikha Subash}, \bibinfo{person}{Alireza Golgouneh}, \bibinfo{person}{Brad Holschuh}, {and} \bibinfo{person}{Lucy~E. Dunne}.} \bibinfo{year}{2023}\natexlab{}.
\newblock \showarticletitle{Plug-and-{Play} {Wearables}: {A} {Repositionable} {E}-{Textile} {Garment} {System} to {Support} {Custom} {Fit} for {Lower}-{Limb} {Rehabilitation} {Applications}}. In \bibinfo{booktitle}{\emph{Adjunct {Proceedings} of the 2023 {ACM} {International} {Joint} {Conference} on {Pervasive} and {Ubiquitous} {Computing} \& the 2023 {ACM} {International} {Symposium} on {Wearable} {Computing}}} \emph{(\bibinfo{series}{{UbiComp}/{ISWC} '23 {Adjunct}})}. \bibinfo{publisher}{Association for Computing Machinery}, \bibinfo{address}{New York, NY, USA}, \bibinfo{pages}{304--309}.
\newblock
\showISBNx{9798400702006}
\urldef\tempurl%
\url{https://doi.org/10.1145/3594739.3610785}
\showDOI{\tempurl}


\bibitem[Yokota et~al\mbox{.}(2015)]%
        {yokota2015eskin}
\bibfield{author}{\bibinfo{person}{Tomoyuki Yokota}, \bibinfo{person}{Yusuke Inoue}, \bibinfo{person}{Yuki Terakawa}, \bibinfo{person}{Jonathan Reeder}, \bibinfo{person}{Martin Kaltenbrunner}, \bibinfo{person}{Taylor Ware}, \bibinfo{person}{Kejia Yang}, \bibinfo{person}{Kunihiko Mabuchi}, \bibinfo{person}{Tomohiro Murakawa}, \bibinfo{person}{Masaki Sekino}, {et~al\mbox{.}}} \bibinfo{year}{2015}\natexlab{}.
\newblock \showarticletitle{Ultraflexible, large-area, physiological temperature sensors for multipoint measurements}.
\newblock \bibinfo{journal}{\emph{Proceedings of the National Academy of Sciences}} \bibinfo{volume}{112}, \bibinfo{number}{47} (\bibinfo{year}{2015}), \bibinfo{pages}{14533--14538}.
\newblock


\end{thebibliography}

\end{document}